\documentclass{article}

\usepackage{PRIMEarxiv}

\usepackage[utf8]{inputenc} 
\usepackage[T1]{fontenc}    
\usepackage{hyperref}       
\usepackage{url}            
\usepackage{booktabs}       
\usepackage{amsfonts}       
\usepackage{nicefrac}       
\usepackage{microtype}      
\usepackage{lipsum}
\usepackage{fancyhdr}       
\usepackage{graphicx}       
\graphicspath{{media/}}     

\usepackage{multirow}
\usepackage{bm}
\usepackage{combelow}
\usepackage[numbers]{natbib}
\usepackage{float}
\usepackage{subfig} 
\usepackage{amsmath}
\usepackage{mathtools}
\usepackage{enumitem}
\usepackage[ruled,vlined,linesnumbered,noend]{algorithm2e}

\title{ContCommRTD: A Distributed Content-based Misinformation-aware Community Detection System for Real-Time Disaster Reporting}

\author{
  Elena-Simona Apostol$^1$, Ciprian-Octavian Truic{\u{a}}$^1$, Adrian Paschke$^2$ \\
  $^1$University Politehnica of Bucharest, Bucharest, Romania \\
  $^2$Fraunhofer Institute for Open Communication Systems, Berlin, Germany \\
  \texttt{alex.petrescu@upb.ro, ciprian.truica@upb.ro, elena.apostol@upb.ro} \\
  \texttt{adrian.paschke@fokus.fraunhofer.de}
}

\begin{document}
\maketitle

\begin{abstract}
Real-time social media data can provide useful information on evolving hazards.
Alongside traditional methods of disaster detection, the integration of social media data can considerably enhance disaster management.
In this paper, we investigate the problem of detecting geolocation-content communities on Twitter and propose a novel distributed system that provides in near real-time information on hazard-related events and their evolution.
We show that content-based community analysis leads to better and faster dissemination of reports on hazards.
Our distributed disaster reporting system analyzes the social relationship among worldwide geolocated tweets, and applies topic modeling to group tweets by topics.
Considering for each tweet the following information: user, timestamp, geolocation, retweets, and replies, we create a publisher-subscriber distribution model for topics.
We use content similarity and the proximity of nodes to create a new model for geolocation-content based communities.
Users can subscribe to different topics in specific geographical areas or worldwide and receive real-time reports regarding these topics.
As misinformation can lead to increase damage if propagated in hazards related tweets, we propose a new deep learning model to detect fake news.
The misinformed tweets are then removed from display.
We also show empirically the scalability capabilities of the proposed system.
\end{abstract}


\maketitle

\section{Introduction}\label{sec:introduction}

Worldwide, hazards frequently have a dramatic impact on rural and urban societies, or the environment.
There are different classes of hazards, e.g., natural (geophysical, hydrological, etc.), anthropogenic (chemical, major accident, technological, biological, etc.).
In recent years many studies have recognized the value of community activity during disasters due to hazards~\cite{Walther2013,Starkey2017}.
Following these analyses, it was observed that people who are physically nearby a location where a hazard occurred tend to produce more disaster-related information on social media~\cite{Cervone2015,Li2017}.

Alongside traditional methods of hazard detection, e.g., satellite or airborne sensing platforms, topographic data sources, Internet of Things, the integration of social media data would considerably enhance disaster management, especially in areas with little to no infrastructure, and, therefore, will be of greater importance in the future~\cite{Smith2015,Rosser2017}.
Twitter is one of the largest social networking platforms, and it has a remarkable capability of continuous retrieval of data and content-sharing services~\cite{Guille2015}.
It is also a suitable place where citizens can present their concerns in a real-time manner~\cite{Fohringer2015}.

Social media has a significant impact on public perception of different hazards and their contingency plans.
This is why is extremely important not to consider tweets that contain misinformation.
Although several papers examined the negative impact of misinformation spread on Twitter during disasters that can cause mass panic or financial loss~\cite{Wang2018rumor}, in the current literature there is no work that mitigates their harm in real time.

In this study, we attempt to answer a fundamental question: Can the social and geographical disparities of Twitter be used to assist citizens and government organizations in informing or disaster management?
To answer this question, we consider classifying tweets in different disaster-related topics and construct a graph model based on content and geolocation.
We also recognise the ethical issues caused by the fact that tweets collected from unreliable sources and showed to other users may increase the spread of misinformation~\cite{Truica2022}. Thus, we also apply fake news detection before displaying the data to the end user.

In this paper, we 
1) investigate the problem of detecting content-based communities in order to provide valuable information on hazard-related events and their evolution, and 
2) propose a distributed content-based misinformation-aware community detection system for real-time disaster reporting (ContCommRTD).
Besides the proposed system, the main contributions are:
\begin{itemize}
    \item[\textit{1)}] We propose a new distributed solution to analyze the social relationship among worldwide geolocated tweets.
    \item[\textit{2)}] We define a new social network model for detecting geolocation-content based communities and named entities.
    \item[\textit{3)}] We create a meta-knowledge dictionary to collect topic-specific tweets containing hazard-related keywords and hashtags.
    \item[\textit{4)}] We apply topic modeling to discover topics and to classify tweets into different topics. A user can then subscribe to one of these topics and receive in real-time all the tweets that discuss the event of interest.
    \item[\textit{5)}] We propose an efficient distributed publish-subscribe based model using Apache Kafka~\cite{Garg2013,Hesse2020} and MongoDB~\cite{Chodorow2010} that allows real-time collection, storage, and distribution of Twitter posts relevant to the chosen type of hazard.
    \item[\textit{6)}] We propose a new Deep Learning architecture, FN-BERT-TFIDF, to determine if the tweets of interest are fake news.
\end{itemize}

The rest of this paper is structured as follows: In Section~\ref{sec:rw}, we present the state-of-the-art. In Section~\ref{sec:method}, we describe the proposed system. In Section~\ref{sec:eval}, we present the datasets and analyze our results. Section~\ref{sec:conclusions} concludes the paper and hints at future research.

\section{Related Work}\label{sec:rw}

Real-time disaster detection using data from dynamic social media environments, e.g., Twitter, and considering different locations is a challenging task that recently received much attention~\cite{Hamoui2020,HernandezSuarez2019,Lukasiewicz2014}.
\cite{Ferner2020} propose detecting disaster-related topics on Twitter data using an LDA (Latent Dirichlet Allocation) topic model enhanced with a set of seed words from older Tweets of the same geographic area. 
This solution is applicable when there is a single topic of interest, e.g., earthquakes or hurricanes, and similar data is available.
Several solutions use in the detection task, not only the textual data from tweets but also incorporate images from the posts or satellites~\cite{Said2019,Huang2019,Jony2019}.
\cite{Arachie2020} present an unsupervised learning solution that detects large-scale hazard-related sub-events in Tweets.
The authors use an ontology containing crisis management vocabulary to rank the candidate sub-events and then cluster the most important sub-events using spectral clustering.

Geolocation is also important when considering natural hazards analysis using social media data. 
To the best of our knowledge, 
the current real-time unsupervised disaster detection solutions consider only geotagged tweets~\cite{Zhang2017,Zhang2018}.
However, tweets do not necessarily come with geoinformation.
Several offline classifier-based disaster detection solutions also try to determine the location by searching the tweets' content for names of cities and countries~\cite{Loynes2020,Suwaileh2020}.
These solutions use corpora of labeled tweets.
\cite{Loynes2020} propose using GeoText to search for names of cities and countries.
Another solution proposes location mention recognition from labeled crisis-related tweets using BERT (Bidirectional Encoder Representations from Transformers~\cite{Devlin2019}) based classification~\cite{Suwaileh2020}.

In the current literature, different graph-based methods are applied to citizen science-based disaster detection solutions.
\citet{Dou2021} use a semantic graph-based topic detection method to 
identify fine-grained topics during natural disasters in social media.
A community detection algorithm, i.e., the algorithm of modularity optimization~\cite{Clauset2004}, is also used to extract topics that denote the same class of event information.
A similar pipeline is also proposed in article~\cite{Ma2019}:
\textit{1)} a graph generation algorithm is used to transform the text data into a graph of the keywords,
\textit{2)} a community detection algorithm is applied to discover hazard-related topics.
\citet{Nguyen2021} present a framework for
\textit{1)} filtering and classifying tweets and
\textit{2)} identifying and summarizing important disaster-related topics. 
The authors use a graph-based ranking algorithm to select and summarize important tweets.

Another issue to consider regarding this research topic is the impact of misinformation.
Misinformation (false or inaccurate information) and disinformation (intentionally spreading misinformation) in the form of fake news are tools used to manipulate public opinion on particular topics, distort public perceptions, and generate social unrest while lacking the rigor of traditional journalism~\cite{Ilie2021,Truica2022}.
\citet{Singh2020} propose content and context aware RNN-based solution for fake news detection during natural disasters. Their solution uses the user profile information and the temporal and textual features of the analysed events.
\citet{Pelrine2021} analyze the performance for misinformation detection of several transformer language models (e.g., BERT~\cite{Devlin2019}, RoBERTa~\cite{Liu2019}, ALBERT~\cite{Lan2020}) on different datasets.
Based on their results, these transformers models have very good performance metrics for large-enough datasets. However, if we consider smaller datasets, such as the case of disaster-related Twitter datasets (e.g., COAID~\cite{Cui2020}), the performance decreases.
Although several studies have analysed the impact of misinformation on different types of hazards, e.g., Hurricane~\cite{Hunt2020misinformation}, COVID-19~\cite{Forati2021geospatial}, none offer a real-time hazard-related event detection solution while removing misinformation.

\section{Problem definition}~\label{sec:problem}

We define an undirected graph $G = (V, E, C, L)$, where $V$ is the set of vertices, $E$ is the set of edges, $C$ is the set that stores the textual content of each vertex, and $L$ the set of geolocations for each node. 
Thus, for our problem, we define the social network as $n$ undirected graphs $\Gamma = \{ G_i = (V_i, E_i, C_i, L_i) |  u, v \in V_i \wedge (u, v) \in E_i \wedge c_u, c_v \in C_i \wedge l_u, l_u \in L_i \wedge i=\overline{1, N}\}$, where each two vertices $u, v \in V_i$ are linked by through an edge $(u,v) \in E_i$ that represents the social relationship between them. 
Thus, $\Gamma$ is a graph with $n$ disconnected components $G_i$.
The content $c_u, c_v$ of vertices $u, v$ is stored in $C_i$, while their geolocation $l_u, l_v$ are stored in $L_i$.
We define a social relationship as the linked that occurs between these nodes in either of the cases: 
\begin{itemize}
    \item[\textit{1)}] $u$ retweets the content of $v$, 
    \item[\textit{2)}] $u$ replies to $v$.
\end{itemize}

For each hazard related event, we have a Meta-Knowledge dictionary that contains a list $H$ with the most relevant keywords and hashtags.
The dictionary is constructed separately. 
Using $H$, we collect the data from the social network.
For each record $u$, we verify if the location $l_u$ is given.
If $l_u$ is missing, we use Name Entity Recognition ($NER$) to extract any mention of locations and then map the name with its coordinates (geolocation).
Using these information together with the social relationships, we construct the list of $n$ undirected graphs $\Gamma$, i.e., the social graph.

Algorithm~\ref{alg:socialgraph} presents the collection and creation of the social graph $\Gamma$, which receives as input the list of relevant keywords and hashtags for an event $H$.
Lines~\ref{line1_sg} to~\ref{line2_sg} initializes the social graph $\Gamma$ and the disconnected graph $\Omega$ used for extracting the connected components $G_i$.
Lines~\ref{line3_sg} to~\ref{line6_sg} initializes the components of $\Omega$
The records are collected into $R$ using a Social Network API (Line~\ref{line7_sg}).
For each record $u \in R$, we add elements to the $V, C, L$ components of $\Omega$ (Lines~\ref{line8_sg} to~\ref{line15_sg}).
Line~\ref{line10_sg} applies some preprocessing techniques on the textual content.
Lines~\ref{line12_sg} to~\ref{line14_sg} determines the coordinates based on locations if the geolocation is not provided in the record.
The edges are added separately after we have all the nodes (Lines~\ref{line16_sg} to ~\ref{line19_sg}).
We construct $\Omega$ (Line~\ref{line20_sg}), extract the connected components to build $\Gamma$ (Line~\ref{line21_sg}), and return the social graph (Line~\ref{line22_sg}).

\begin{algorithm}[!ht]
\DontPrintSemicolon

\SetKwInOut{Input}{Input}
\SetKwInOut{Output}{Output}
\Input{the keywords and hashtags list $H$}
\Output{the undirected social graph $\Gamma$}

\emph{$\Gamma \gets \emptyset$}\;\label{line1_sg}
\emph{$\Omega \gets \emptyset $}\;\label{line2_sg}
\emph{$V \gets \emptyset$}\;\label{line3_sg}
\emph{$E \gets \emptyset$}\;\label{line4_sg}
\emph{$C \gets \emptyset$}\;\label{line5_sg}
\emph{$L \gets \emptyset$}\;\label{line6_sg}

\emph{$R \gets SocialNetworkAPI(H)$}\;\label{line7_sg}

\ForEach{$u \in R$}{\label{line8_sg}
    \emph{$V \gets V \cup \{u \}$}\;\label{line9_sg}
    \emph{$c'_u \gets Preprocess(c_u)$}\;\label{line10_sg}
    \emph{$C \gets C \cup \{c'_u\}$}\;\label{line11_sg}
    \If{$l_u = Nil$}{\label{line12_sg}
        \emph{$\lambda \gets NER(c_u)$}\;\label{line13_sg}
        \emph{$l_u \gets GeoLocation(\lambda)$}\;\label{line14_sg}
    }
    \emph{$L \gets L \cup \{l_u\}$}\;\label{line15_sg}
}

\ForEach{$u \in V$}{\label{line16_sg}
    \ForEach{$v \in V$}{\label{line17_sg}
        \If{$Retweet(u, v) \vee Reply(u, v)$}{\label{line18_sg}
            \emph{$E \gets E \cup \{(u, v) \}$}\;\label{line19_sg}
        }
    }
}

\emph{$\Omega = (V, E, C, L)$}\;\label{line20_sg}
\emph{$\Gamma \gets ConnectedComponents(\Omega)$}\;\label{line21_sg}

\Return{$\Gamma$}\;\label{line22_sg}
\caption{$SocialGraph$ - Construct the social graph}
\label{alg:socialgraph}
\end{algorithm}

Using the graph structures in $\Gamma$, we want to determine communities where there is a high social media activity for users' topics of interest.
These communities can contain multiple graphs $G_i \in \Gamma$ that are not directly interconnected.
To build the communities, first we filter the content of a graph $G_i$ and remove any nodes and their edges that spread misinformation.
At the end of the misinformation detection process we obtaining $\Gamma'$ that contain only the clean undirected graphs $G'_i$.

Algorithm~\ref{alg:misinformation} presents the misinformation detection algorithm, which receives as input the social graphs $\Gamma$ and outputs the clean graph $\Gamma'$.
After the clean graph is initialized $\Gamma'$ (Line~\ref{line1_md}), the veracity of all the nodes' content from a graph $G \in \Gamma$ is verified (Lines~\ref{line2_md} to~\ref{line15_md}).
If the the content is deemed as Fake, the record is removed form the graph (Lines~\ref{line7_md} to~\ref{line13_md}.
The corresponding graph $G'_i$ is constructed using the updated $V', E', C', L'$ and it is added to $\Gamma'$ (Lines~\ref{line14_md} to~\ref{line15_md}).
When the verification are finished for all the graphs, the algorithm returns the clean graph $\Gamma'$.

\begin{algorithm}[!htbp]
\DontPrintSemicolon

\SetKwInOut{Input}{Input}
\SetKwInOut{Output}{Output}
\Input{the social graphs $\Gamma$}
\Output{the clean undirected graph $\Gamma'$}

\emph{$\Gamma' \gets \emptyset$}\;\label{line1_md}

% ...
\ForEach{$G = (V, E, C, L) \in \Gamma$}{\label{line2_md}
    \emph{$C' \gets C$}\;\label{line5_md}
    \ForEach{ $c_u \in C$ }{\label{line7_md}
        \emph{$veracity \gets DetectVeracity(c_u)$}\;\label{line8_md}
        \If{$ veracity = Fake $}{\label{line9_md}
            \emph{$V \gets V \setminus \{u \}$}\;\label{line10_md}
            \emph{$E \gets E \setminus \{(u, v) | (u, v) \in E \}$}\;\label{line11_md}
            \emph{$C' \gets C' \setminus \{c_u\}$}\;\label{line12_md}
            \emph{$L \gets L \setminus \{l_u\}$}\;\label{line13_md}
        }
    }
    \emph{$G' = (V, E, C', L)$}\;\label{line14_md}
    \emph{$\Gamma' = \Gamma' \cup \{ G \} $}\;\label{line15_md}
}

\Return{$\Gamma' $}\;\label{line22_md}
\caption{$MisinformationDetection$ - Content veracity detection}
\label{alg:misinformation}
\end{algorithm}

With the clean content $\Gamma'$, we build the content-based communities. 
In order to achieve this, we utilize:
\begin{itemize}
    \item[\textit{1)}] the content similarity between the nodes given the membership level $m$ of the content of a node $c_u$ to a topic $t_k$, and
    \item[\textit{2)}] the proximity of nodes to a core point $p$ that defines a geographic area $A$.
\end{itemize}

To compute the content similarity, we first use a topic modeling algorithm to extract $k$ topics $T$ within our graphs. 
Then we determine the membership of a node's content $c_u$ to belong to a topic $t_j \in T$ ($j=\overline{1, k}$) using the cosine similarity $sim(c_u, t_j)$.
Given the similarity $sim(c_u, t_j)$ between a node's content and a topic, if the similarity is over a given threshold $\varepsilon_c$ ($sim(c_u, t_j) \leq \varepsilon_c$) then $c_u$ belongs to topic $t_j$.
Based on the threshold, the same node's content can belong to multiple topics.
Using the topics $T$ and the graph $\Gamma'$, we construct the topic graphs $\Theta$.

Algorithm~\ref{alg:topicgraph} presents the topic graphs extraction. 
The algorithm receives as input the the undirected clean graphs $\Gamma'$, the number of topics $k$, and the similarity threshold $\varepsilon_c$. 
The output is the topic graphs $\Theta$.
Line~\ref{line1_tm} initializes $\Theta$.
Lines~\ref{line2_tm} to~\ref{line5_tm} concatenates the contenct of all the graphs $G$ and extract $k$ topics stored in $T$.
We iterate trough each graph and each topic to determine the membership of the content of a node to a topic using the similarity and create a topic graph $G'$ which is added to the topic graphs $\Theta$ (Lines~\ref{line6_tm} to~\ref{line19_tm}).
At the end of the iteration, the topic graphs $\Theta$ are returned (Line~\ref{line20_tm}).

\begin{algorithm}[!ht]
\DontPrintSemicolon
\SetKwInOut{Input}{Input}
\SetKwInOut{Output}{Output}
\Input{the undirected clean graphs $\Gamma'$ \newline 
    the number of topics $k$ \newline
    the similarity threshold $\varepsilon_c$
}
\Output{the topic graphs $\Theta$}

\emph{$\Theta \gets \emptyset$}\;\label{line1_tm}

\emph{$D \gets \emptyset $}\;\label{line2_tm}
\ForEach{$G = (V, E, C, L) \in \Gamma'$}{\label{line3_tm}
    \emph{$D \gets D \cup C $}\;\label{line4_tm}
}

\emph{$T \gets Topics(D, k) $}\;\label{line5_tm}

\ForEach{$ t \in T$}{\label{line6_tm}
    \emph{$V' \gets \emptyset$}\;\label{line7_tm}
    \emph{$E' \gets \emptyset$}\;\label{line8_tm}
    \emph{$C' \gets \emptyset$}\;\label{line9_tm}
    \emph{$L' \gets \emptyset$}\;\label{line10_tm}
    \ForEach{$G = (V, E, C, L) \in \Gamma'$}{\label{line11_tm}
        \ForEach{$ c_u \in C$}{\label{line12_tm}
            \If{$ sim(c_u, t) \geq \varepsilon_c $}{\label{line13_tm}
                \emph{$V' \gets V' \cup \{ u \}$}\;\label{line14_tm}
                \emph{$E' \gets E' \cup \{ (u,v) | (u,v) \in E \}$}\;\label{line15_tm}
                \emph{$C' \gets C' \cup \{ c_u \} $}\;\label{line16_tm}
                \emph{$L' \gets L' \cup \{ l_u \} $}\;\label{line17_tm}
            }
        }
    }
    \emph{$G' = (V', E', C', L')$}\;\label{line18_tm}
    \emph{$\Theta \gets \Theta \cup \{ G' \}$}\;\label{line19_tm}
}

\Return{$\Theta $}\;\label{line20_tm}
\caption{$TopicGraphs$ - Topic graphs extraction}
\label{alg:topicgraph}
\end{algorithm}

The proximity a node $u$ to a core point $p$ is computed using its geolocation coordinates $l_u$.
The intuition behind this assumption is based on the fact that nodes that are near each other will fall within the same geographic area $A$.
Each geographic area $A$ represents a cluster.
The area $A$ is given by a core point $p$ and a radius $\varepsilon_l$ that specifies the cluster's maximum extend.
Thus, the proximity of a node $u$ is computed as distance $\delta(l_u, p)$ between its geolocation $l_u$ and the core point $p$.
If $\delta(l_u, p) \leq \varepsilon_l$ then $u \in A$.
The area $A$ and the core point $p$ is determined using a data clustering algorithm, e.g., DBSCAN~\cite{Ester1996}.
Using the content similarity and the proximity of nodes, we create geolocation-content based communities $\Sigma = (V_{\Sigma}, E_{\Sigma}, C_{\Sigma}, L_{\Sigma})$.
Within the same area, there can be multiple communities.

Algorithm~\ref{alg:geocom} presents the construction of the geolocation-content based communities.
The algorithm receives as input the topic graph $\theta \in \Theta$ and the proximity threshold $\varepsilon_l$ and outputs the communities $\Sigma$.
Line~\ref{line1_com} initializes $\Sigma$.
Lines~\ref{line2_com} to~\ref{line5_com} extract all the geolocations from $\theta$ and determines the areas $A$ for which the radius is equal to $\varepsilon_l$.
We iterate trough each graph and each area to construct the communities $\Sigma$ using the condition $\delta(l_u, p) \leq \varepsilon_l$  (Lines~\ref{line6_com} to~\ref{line19_com}).
At the end of the iteration, the geolocation-content based communities $\Sigma$ are returned (Line~\ref{line20_com}).

\begin{algorithm}[!ht]
\DontPrintSemicolon

\SetKwInOut{Input}{Input}
\SetKwInOut{Output}{Output}
\Input{the undirected topic graph $\theta$ \newline 
    the proximity threshold $\varepsilon_l$ 
}
\Output{the communities  $\Sigma' = (V_{\Sigma}, E_{\Sigma}, C_{\Sigma}, L_{\Sigma})$}

\emph{$\Sigma' \gets \emptyset$}\;\label{line1_com}

\emph{$P \gets \emptyset $}\;\label{line2_com}
\ForEach{$(V, E, C, L) \in \theta$}{\label{line3_com}
    \emph{$P \gets P \cup L $}\;\label{line4_com}
}

\emph{$A \gets Areas(P, \varepsilon_l) $}\;\label{line5_com}

\ForEach{$p \in A$}{\label{line6_com}
    \emph{$V' \gets \emptyset$}\;\label{line7_com}
    \emph{$E' \gets \emptyset$}\;\label{line8_com}
    \emph{$C' \gets \emptyset$}\;\label{line9_com}
    \emph{$L' \gets \emptyset$}\;\label{line10_com}
    \ForEach{$(V, E, C, L) \in \theta$}{\label{line11_com}
        \ForEach{$ l_u \in L$}{\label{line12_com}
            \If{$ \delta(l_u, p) \leq \varepsilon_l $}{\label{line13_com}
                \emph{$V' \gets V' \cup \{ u \}$}\;\label{line14_com}
                \emph{$E' \gets E' \cup \{ (u,v) | (u,v) \in E \}$}\;\label{line15_com}
                \emph{$C' \gets C' \cup \{ c_u \} $}\;\label{line16_com}
                \emph{$L' \gets L' \cup \{ l_u \} $}\;\label{line17_com}
            }
        }
    }
    \emph{$G' = (V', E', C', L')$}\;\label{line18_com}
    \emph{$\Sigma' \gets \Sigma' \cup \{ G' \}$}\;\label{line19_com}
}

\Return{$\Sigma'$}\;\label{line20_com}
\caption{$CommunityGraphs$ - Geolocation-content based communities extraction}
\label{alg:geocom}
\end{algorithm}

Algorithm~\ref{alg:ContCommRTD} presents ContCommRTD, the solution to our problem.
The algorithm receives as input the the keywords and hashtags list $H$, the number of topics $k$, the similarity threshold $\varepsilon_c$, and the proximity threshold $\varepsilon_l$.
Firstly, the social graph $\Gamma$ is build using $H$ and Algorithm~\ref{alg:socialgraph} (Line~\ref{line1}).
Secondly, the graph $\Gamma'$ is constructed by removing from $\Gamma$ all the nodes that contain misinformation (Line~\ref{line2}).
Thirdly, the topic graphs are extracted (Line~\ref{line3}).
Lastly, the geolocation-content based communities $\Sigma$ are determined for each topic graph and returned (Lines~\ref{line4} to~\ref{line6}).

\begin{algorithm}[!ht]
\DontPrintSemicolon

\SetKwInOut{Input}{Input}
\SetKwInOut{Output}{Output}
\Input{ the keywords and hashtags list $H$ \newline 
    the number of topics $k$ \newline
    the similarity threshold $\varepsilon_c$ \newline 
    the proximity threshold $\varepsilon_l$ 
}
\Output{the communities  $\Sigma = (V_{\Sigma}, E_{\Sigma}, C_{\Sigma}, L_{\Sigma})$}

\emph{$\Sigma \gets \emptyset$}\;\label{line0}
\emph{$\Gamma \gets SocialGraph(H)$}\;\label{line1}
\emph{$\Gamma' \gets MisinformationDetection(\Gamma)$}\;\label{line2}
\emph{$\Theta \gets TopicGraphes(\Gamma', k, \varepsilon_c)$}\;\label{line3}
\ForEach{$G = (V, E, C, L) \in \Theta$}{\label{line4}
    \emph{$\Sigma \gets \Sigma \cup CommunityGraphs(G, \varepsilon_l)$}\;\label{line5}
}

\Return{$\Sigma$}\;\label{line6}
\caption{\textit{ContCommRTD} - Geolocation  community graphs extraction}
\label{alg:ContCommRTD}
\end{algorithm}

\section{System Description}\label{sec:method}

\begin{figure*}[!htbp]
\centering
    \includegraphics[width=1\textwidth]{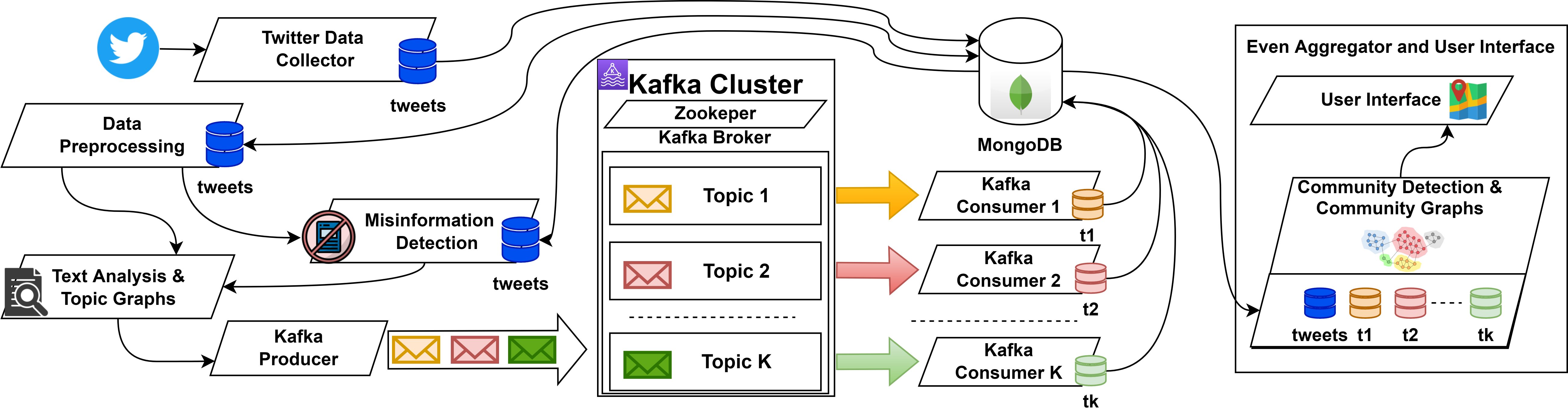}
    \caption{Architecture of \textit{ContCommRTD} (database colors denote the collections used by each module)}
    \label{fig:architecture}
\end{figure*}

The architecture of the proposed system \textit{ContCommRTD}, which implements Algorithm~\ref{alg:ContCommRTD}, is presented in Figure~\ref{fig:architecture}.

\subsection{Twitter Data Collector}

This module is used to fetch Twitter data in real-time, using the Twitter Developer API and it implements Line~\ref{line7_sg} from Algorithm~\ref{alg:socialgraph}.
It collects tweets based on the chosen type of hazard, e.g., extreme hydrological hazards associated with water-related events.
The initial selection is done by searching tweets that contain specific words. 
For this purpose, we created a meta-knowledge dictionary as follows.
For each new type of hazard, we add a new entry containing a list of top-$k$ keywords and hashtags for the chosen hazardous event.
To extract the keywords we employ TLATR~\cite{Truica2021TLATR} using the following pipeline: 
\begin{itemize}
    \item[\textit{1)}] extract the topics and keywords including hashtags 
    \item[\textit{2)}] label the topics. 
\end{itemize}

When users searcher for new events, they  will select from a list of topics' labels and the system will automatically load all the topic's keywords and hashtags, which in turn are used to filter the stream of tweets.
The collected tweets are stored in \textit{tweets} collection in a NoSQL Document-Oriented Distributed Database Management system, i.e., MongoDB.
We choose MongoDB, because benchmarks show that this system is fast, reliable, and offers good performances when dealing with textual data~\cite{Truica2020,Truica2021a}.

Using these filters, we collect the tweet and its retweets and replies.
For a tweet, we store the following information: id, language, retweet flag, creation date, user information, geolocation information, and text.
For a retweet, we also keep the retweeter's coordinates and location.
Each retweet information is also updated in real-time with the original tweet's data. 
The same information is also kept for replies.
We use geolocation as a sharding key to distribute the data between different sites and improve querying.

\subsection{Data Preprocessing and Geolocation enhancement}

This module implements constructs the social graph $\Gamma$ implemented by Algorithm~\ref{alg:socialgraph}.

\subsubsection{Data Preprocessing.}
We extract the textual content of all the Tweets from the database and apply the following preprocessing steps to extract a \textit{clean text}:
\begin{itemize}
    \item[\textit{1)}] remove user mentions (terms that start with @), http links, punctuation, double spacing, and numbers from the text;
    \item[\textit{2)}] Transform the text to lowercase;
    \item[\textit{3)}] Extract terms through tokenization;
    \item[\textit{4)}] Remove stop words and tokens with a length smaller than 3 characters;
    \item[\textit{5)}] Extract the stem of the remaining terms using the Porter Stemmer~\cite{Porter1980} to reduce the number of terms by removing inflected and derived words, thus minimizing the vocabulary.
\end{itemize}

Using the clean text and the normalized Term Frequency-Inverse Document Frequency $TFIDF_{n}$, we built the document-term matrix $W=[w_{ij}]$ needed by the analysis algorithm, where $w_{ij}=TFIDF_{n}(t_{ij},d_i,D)$, $t_{ij}$ is a term appearing in document $d_{i}$, and $D$ is the entire collection of tweets.
We use the \textit{scikit-learn}~\cite{Pedregosa2011} $TFIDF$ implementation.

\subsubsection{Analyzing the Geolocation of Tweets}

We utilize several techniques to obtain the geolocation information of a tweet, retweet, or reply.
The most basic approach was to get the coordinates of a tweet. 
For this to work, the user must have the location turned on.
There are various studies (e.g., \cite{snyder2019city}) that show that only a relatively small percentage of users use this feature.
However, we observe that a large majority of the collected hazard-related tweets are either geotagged or contain location information in the text, e.g., "Heavy rain in Masjid Al Haram".
A tweeter that contains a Place has 4 pairs of coordinates that define the area for that Place.
If the tweet doesn't have a geolocation tag either, we apply Named Entity Recognition to identify possible locations from the text.
Once identified, we use GeoNames~\cite{geonames} to get the coordinates.
This is an improvement over many existing real-time systems that solely rely on geolocation information embedded within the tweet to determine events~\cite{Zhang2017,Zhang2018}.

\subsection{Misinformation Detection}

\begin{figure}[!htb]
    \centering
    \subfloat[FN-BERT-TFIDF\label{fig:fn_bert_tfidf}]{{\includegraphics[width=\columnwidth]{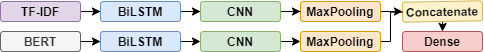}}}
    \hfill    
    \subfloat[FN-TFIDF\label{fig:fn_tfidf}]{{\includegraphics[width=\columnwidth]{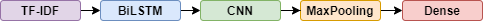}}}
    \hfill
    \subfloat[FN-BERT\label{fig:fn_bert}]{{\includegraphics[width=\columnwidth]{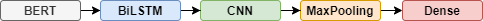}}}

    \caption{Deep Learning Architectures}
\end{figure}

This module is used to filter and mark any tweets that spread misinformation, i.e., Fake News (FN).
This module implements Algorithm~\ref{alg:misinformation}.
As the Twitter datasets that contain disaster-related events have usually a limited size, we use Transfer Learning techniques to transfer knowledge gained from these larger but more generic Twitter datasets to misinformation detection on disaster-related datasets.
For the task of misinformation detection, we propose a new Deep Learning architecture, FN-BERT-TFIDF (Figure~\ref{fig:fn_bert_tfidf}), that receives as input the TF-IDF vector for a tweet as well as tweet embedding extracted with BERT.
For each input, the model contains one BiLSTM layer, a CNN layer, and a MaxPooling layer.
We choose a BiLSTM layer because it enables the network to use both previous and future elements by looking forward and backward in the words sequence.
We use a CNN layer to create new features using the convolution operation between the text window and every distinct filter.
We use a MaxPooling layer to decrease the size of the feature channels by grouping elements into fix lengths sequences and choosing only the feature with the maximum value.
The output of the MaxPooling layers are concatenated and sent as input to the classification layer.
To test the efficiency of our model, we perform ablation testing and compare the results of the FN-TFIDF-BERT model with the results obtained when only TF-IDF, i.e., FN-TFIDF (Figure~\ref{fig:fn_tfidf}), or BERT, i.e., FN-BERT (Figure~\ref{fig:fn_bert}), is used as input.
After a tweet passes throughout the detection model, its record in MongoDB is updated with a new field that encodes its veracity.
If a tweet is detected as fake, then it will not be used in the next modules.
For implementing the Deep Learning models, we use the \textit{Keras} interface of the \textit{TensorFlow}~\cite{tensorflow2015}.
For building the BERT embeddings, we use \textit{simpletransformers} with the \textit{HuggingFace}~\cite{Wolf2020} BERT model.

\subsection{Text Analysis and Topic Graphs}

This module implements Algorithm~\ref{alg:topicgraph} to extract the topic graphs.
To Analyze the tweets and determine the context regarding the hazard discussed in the text, we use Online Latent Dirichlet Allocation (OLDA)~\cite{AlSumait2008}. 
OLDA is a generative statistical model used for topic modeling that groups together terms that are syntactically different but have similar meaning and represent the same concepts. 
The algorithm determines for each tweet a specific topic by calculating the similarity between the tweet and all of the topics.
Thus, OLDA assigns a tweet to a mixture of topics, i.e., each tweet is a combination of one or more topics.
The model is built in real-time, as more tweets are added to the database, the model is retrained with the new information.
We use a 50\% threshold (i.e., $\varepsilon_c$) to determine if a tweet is relevant to a topic as a tweet can belong to multiple topics.
We use the \textit{gensim}~\cite{Rehurek2010} OLDA implementation.

\subsection{Kafka Modules}

The \textit{Kafka Producer} module takes $<$tweet, topic$>$ pairs and sends them to the \textit{Kafka Consumers}.
\textit{Kafka Cluster} is used to create an environment that assures the distributed exchange of messages between the \textit{Kafka Producer} and the \textit{Kafka Consumers}.
It contains the \textit{Kafka Broker} and the \textit{Zookeeper} service.
The job orchestration and cluster topology is done by the \textit{Zookeeper} service, a distributed coordination service for distributed applications~\cite{Artho2019}.
The \textit{Kafka Broker}'s main role is to act as a transition channel. 
It takes messages in the form of $<$tweet, topic$>$ pairs and delivers them to the right \textit{Kafka Consumer}.
Thus, the receiving of data by the \textit{Kafka Consumers} is done in parallel in real-time in a non-blocking way.
Each \textit{Kafka Consumer} stores messages in their own topic-dependent collections within MongoDB and stores $<$topic, [(tweet\_ids, geolocation)]$>$ pairs, i.e., a topic and a list of all the related tweets ids with their geolocation.
The geolocation is used for sharding.

\subsection{Community Graphs and User interface}

\begin{figure}[!htbp]
    \centering
    \subfloat[Google map view\label{fig:ui_map}]{{\includegraphics[width=0.38\columnwidth]{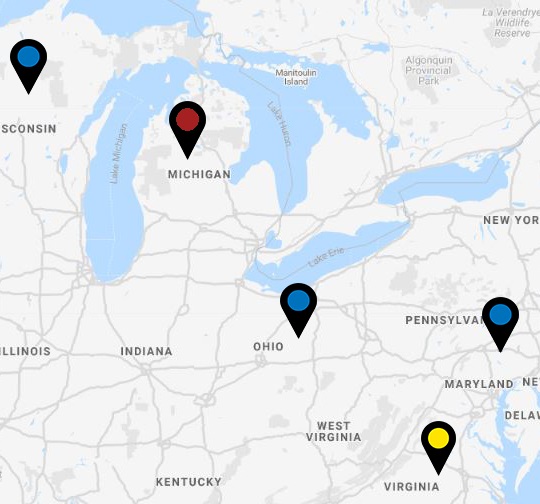}}}
    \hfill    
    \subfloat[Subscriptions\label{fig:ui_subs}]{{\includegraphics[width=0.22\columnwidth]{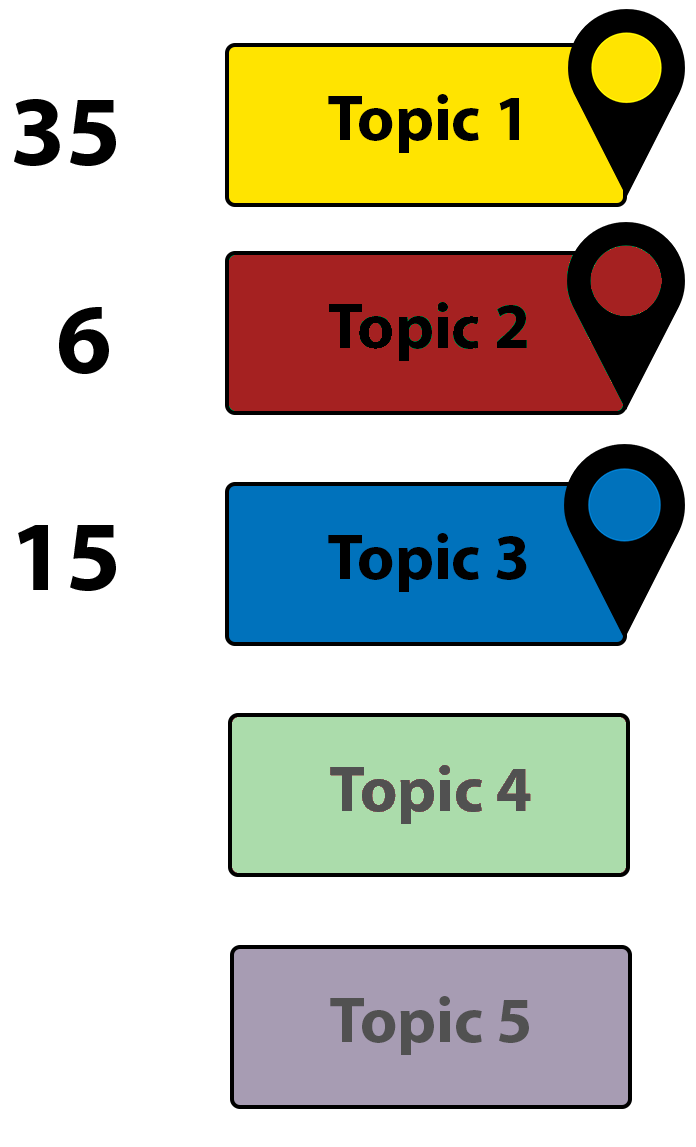}}}
    \hfill    
    \subfloat[Communities per topic\label{fig:ui_c}]{{\includegraphics[width=0.38\columnwidth]{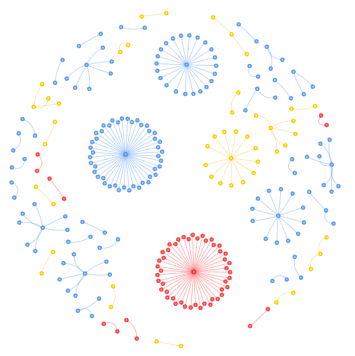}}}
    \caption{User interface}
    \label{fig:ui}
\end{figure}

This module is used to present in real-time to users one or more topics of interest from one or more hazard related fields on Google maps (Figure~\ref{fig:ui_map}).
Users can choose and subscribe to desired topics (Figure~\ref{fig:ui_subs}) and can monitor a specific area on the map just by searching for it.
For each subscribed topic, the user sees on the map a colored pins updated automatically in almost real-time.
A user can also view the formed communities based on the chosen topics (Figure~\ref{fig:ui_c}).

Each pin represent a community for a specific topic within the social graph.
To form these communities, we use Algorithm~\ref{alg:geocom} with DBSCAN (Density Based Spatial Clustering of Applications with Noise)~\cite{Ester1996}, a density-based clustering algorithm.
DBSCAN computes the density of points (core points) $p$ around which we form the area $A$. 
Thus, the clustering algorithm measures density as the number of points within the radius $\varepsilon_l$ of a point from the analyzed dataset.
It can estimate the connected components of the $\lambda$-density level set ${x:f(x)\geq\lambda}$ given $n$ samples from an unknown density $f$~\cite{jiang2017density}.

The user has access to the $<$topic, tweets$>$ pairs.
The topic is extracted from the MongoDB based on the user's subscribed topics.
By matching the tweet\_ids presented in the \textit{tweets} and \textit{topic} collections, we extract all the geographical coordinates and group them together using DBSCAN.
Thus, the same topic can appear multiple times on the map in different geographic locations.
For these geographic locations, we present to the user the latest information as the tweets are updated in real-time by the \textit{Twitter Data Collector} module. 
This component employs strong synchronization, in order to retrieve information from the database in near real-time.

If a user unsubscribes from a topic, then the corresponding pins are removed from the map.
We use timestamped matching queries to retrieve the relevant information.
The pins for the topics that have not beet updated in more than 24h are removed automatically from the map.
Their topics are kept in \textit{tweets} collection for retraining and improving performance of the OLDA model.

\subsection{Design Challenges}

With real-time data stream collection systems, it is important to take into account that the system may be disconnected from the service. 
To mitigate this challenge, we implemented a background process checker that verifies every 30 seconds that the \textit{Twitter Data Collection} is working.
Also, Twitter is limiting the number of tweets that can be collect in an hour.
We address this challenge by using filters to select only relevant tweets.

To build the topic graph, we use OLDA.
The reason behind choosing OLDA, instead of LSI or NMF, is that OLDA is a probabilistic model while NMF and LSI are a matrix factorization and multivariate analysis techniques.
With the increase of the volume of collected data, the runtime performance of OLDA can decrease.
Thus, we systematically start in background a new OLDA training process each hour.
When a training process is finished, the new model is uploaded in the application and the old one is removed from production and archived.
Furthermore, we also use the same approach to fine-tune the misinformation model to improve performance.
We use the same strategy for building the community graph with DBSCAN as we do for constructing the topic graph with OLDA.
We use DBSCAN as we do not need to know apriori the number of clusters.
We use sharding for our database to increase localization, load balancing, and querying through geodistribution. 
For all the collections, we use the geolocation coordinates as the shard key.
We also use replication to create three member replica sets for each shard.
Through this mechanism, we increase data availability, add fault and partition tolerance, and eliminate the single point of failure.
These design choices are also used to seamlessly scale the datasets horizontally.

\section{Evaluation}\label{sec:eval}

In this section, we present the experimental results of our proposed system, ContCommRTD. 
We first focus on the ablation testing for the misinformation detection models.
Secondly, we present two use cases of ContCommRTD for the disaster reporting: 
\begin{itemize}
    \item[\textit{1)}] Hydrological Hazards and 
    \item[\textit{2)}] COVID-19 Infection Hazards.
\end{itemize}

We conclude this section with ContCommRTD scalability tests.

\subsection{Data Collection}

We collect two datasets for the two use cases.
First, for the Hydrological Hazards use case we collect 356\,483 Tweets using the hydrological data dictionary and store them in the distributed MongoDB database.
This dataset is also used to train the first OLDA model and to evaluate the algorithm's runtime performance and topic quality for the first.
Second, for the COVID-19 Infection Hazards we collect 50\,230 and present the community detection results. 
Table~\ref{tab:filtering_kw} presents a subset of the used keywords and hashtags for both hydrological and COVID-19 use cases.

\begin{table}[!htbp]
\caption{Example of filter keywords (subset)}
\label{tab:filtering_kw}
\centering
\begin{tabular}{|l|l|}
\hline
\begin{tabular}[c]{@{}l@{}}\textbf{Keywords} \\ \textbf{flood} \end{tabular} & \begin{tabular}[c]{@{}l@{}}flowage, rain, precipitation, floodplain, groundwater, overflow, deluge\\
water level, water flow, rainfall, inundation, torrent, groundwater flood\\
tsunami, torential, costal flooding, costal storm, river flooding, hurricane, ...
\end{tabular} \\ \hline
\begin{tabular}[c]{@{}l@{}}\textbf{Hashtags} \\ \textbf{flood} \end{tabular} & \begin{tabular}[c]{@{}l@{}}
\#flood, \#precipitation, \#rainfall, \#deluge, \#torrent, \#inundation, \#rain\\
\#floods, \#waterlevel, \#hurricane, \#tornado, \#torential, \#storm, \#flowage, ...
\end{tabular}                                        \\ \hline
\begin{tabular}[c]{@{}l@{}}\textbf{Keywords} \\ \textbf{Covid-19} \end{tabular} & \begin{tabular}[c]{@{}l@{}}
coronavirus, corona, COVID, COVID-19, pandemic, quarantine, lockdown,\\ 
corona virus, hand sanitizer, infection, wash your hands, mask, \\
personal protective equipment, covid quarentined, intensive care ...
\end{tabular} \\ \hline
\begin{tabular}[c]{@{}l@{}}\textbf{Hashtags} \\ \textbf{Covid-19} \end{tabular} & \begin{tabular}[c]{@{}l@{}}
\#covid, \#sarscov2, \#corona, \#vaccine, \#stayhealthy, \#stayathome, \\
\#SARSCoV2 \#viruscorona, \#quarentined, \#quarentinelife, ...
\end{tabular}                                                                                                                                                                                                                   \\ \hline
\end{tabular}
\end{table}

\subsection{Evaluation Metrics}

\subsubsection{Misinformation Detection}

To evaluate the quality of the Deep Learning architectures employed for Misinformation Detection, we use Accuracy, Precision, and Recall.

\subsubsection{Topic Modeling}
To evaluate the quality of the topic model, we employ perplexity and topic coherence~\cite{Newman2010}.
Perplexity is a measure that
determines how well a probability model predicts a sample~\cite{Gupta2020}.
A low perplexity denotes that the distribution predicts correctly the sample. 
For topic modeling, an algorithm that achieves a low perplexity indicates that it fit the data better.
Topic Coherence measures the human-interpretability of a topic~\cite{Lukasiewicz2018}.
We use the $C_V$~\cite{Roder2015} to measure the coherence of our topics.
For perplexity, we use \textit{gensim}'s \textit{log\_perplexity()} implementation, while for $C_V$ we use \textit{palmetto}~\cite{Roder2015}.

\subsubsection{Community Detection}
To evaluate the quality of the communities, we employ Davies-Bouldin~\cite{Davies1979}, Calinski-Harabasz~\cite{Calinski1974}, and Silhouette~\cite{Rousseeuw1987}.
The Davies-Bouldin score measures the separation between clusters by computing the ratio between within-cluster distances and between-cluster distances which determines the average similarity of each cluster with its most similar cluster.
The Davies-Bouldin score bounded in the range $[0, 1]$, whit a score closer to 0 showing better separation. 
The Calinski-Harabasz score measures how well are the clusters defined by computing the ratio between the within-cluster dispersion and the between-cluster dispersion.
A higher Calinski-Harabasz shows core shows clusters that are dense and well separated. Unfortunately, the Calinski-Harabasz score is not bounded.
The Silhouette score is another measurement for determining how well are the clusters defined by computing the mean intra-cluster distance and the mean nearest-cluster distance.
The Silhouette is defined in $[-1, 1]$ range, with scores closer $-1$ for incorrect clusters, scores closer to $+1$ for highly dense clusters, and scores around $0$ for overlapping clusters.

\subsection{Misinformation Detection}

To test the proposed models, we use two publicly available dataset: Covid-19 dataset~\cite{Cheng2021} and LIAR~\cite{Wang2017} using 2 labels, i.e., fake or real. 
For the experiments, we use a 70\%-30\% train-test split with random shuffle, while maintaining the the class ratio for the two sets.
We use 256 units for the BiLSTM with a dropout of 0.2.
For the CNN layer, we use 64 filters and a kernel size of 128.
We used a 5\,000-dimensions TF-IDF vector and the pretrained 1\,024-dimension BERT transformer from \textit{HuggingFace}~\cite{Wolf2020}, i.e., \textit{bert-large-uncased}.
We observed that FN-BERT-TFIDF outperforms the other two models on both dataset (Table~\ref{tab:results_mi}).

\begin{table}[!htbp]
\caption{Misinformation ablation results}
\label{tab:results_mi}
\centering
\begin{tabular}{|c|c|c|c|c|c|}
\hline
\textbf{Dataset} & \textbf{Model} & \textbf{Accuracy} & \textbf{Precision} & \textbf{Recall} & \textbf{F1 Score} \\ \hline
\multirow{4}{*}{LIAR} & FN-TFIDF & 56.01 $\pm$ 0.33 & 55.81 $\pm$ 0.52 & 55.85 $\pm$ 0.54 &55.83 $\pm$ 0.53 \\ \cline{2-6} 
 & FN-BERT & 58.75 $\pm$ 0.44 & 57.97 $\pm$ 0.50 & 57.09 $\pm$ 0.46 & 57.53 $\pm$ 0.48\\ \cline{2-6} 
 & FN-BERT-TFIDF & \textbf{60.92 $\pm$ 0.59} & 60.65 $\pm$ 0.90 & 60.66 $\pm$ 0.97 & 60.65 $\pm$ 0.93\\ \cline{2-6}
 & LSTM~\cite{Rashkin2017} & $\sim$58 & N/A & N/A & N/A \\ \hline
\multirow{4}{*}{Covid-19} & FN-TFIDF & 81.93 $\pm$ 0.37 & 78.48 $\pm$ 0.64 & 76.90 $\pm$ 0.11 &77.68 $\pm$ 0.19 \\ \cline{2-6} 
 & FN-BERT & 86.86 $\pm$ 0.07 & 86.86 $\pm$ 0.07 & 86.86 $\pm$ 0.07 & 86.86 $\pm$ 0.07 \\ \cline{2-6} 
 & FN-BERT-TFIDF & \textbf{87.92 $\pm$ 0.04} & 87.91 $\pm$ 0.02 & 87.92 $\pm$ 0.03 & 87.82 $\pm$ 0.02\\ \cline{2-6}
 & VAE + MLP~\cite{Cheng2021} & N/A & N/A & N/A & 85.98 $\pm$ 0.10  \\ \hline
\end{tabular}
\end{table}

\subsection{Topic and Communities detection}

When training OLDA, we need to initialize two parameters: 
\begin{itemize}
    \item[\textit{1)}] \textit{Alpha} the document-topic density (a larger value means that a document contain a lager number of topics), and
    \item[\textit{2)}] \textit{Beta} the topic-word density (a larger value means that more words are considered to belong to the same topic).
\end{itemize}

We initialize these values to "auto" in order to learn them automatically from the corpus.
After training the OLDA model 10 times on the initial corpus, we obtain an average runtime of 17min.

\subsubsection{Use case 1: Hydrological Hazards.}

We extract 3 topics (Table~\ref{tab:lda_hydro}) and evaluate the quality of the model using perplexity, obtaining a score of $7.83$ for the entire dataset.
This score shows that OLDA manages to predict well the data sample from our dataset.
Finally, we compute the $C_V$ score to determine topic coherence and human-interpretability.
For the OLDA model train on the initial corpus we obtain a $C_V=0.48$, meaning that the topics are readable by users.
The evaluation of the community graphs shows a good separation and well-bounded clusters. The scores we obtain are the following: 0.0286 Davies-Bouldin, $\sim 3.60e9$ Calinski-Harabasz, and 0.9758 Silhouette.

\begin{table}[!htbp]
\caption{Example of OLDA topics for hydrological hazards}
\label{tab:lda_hydro}
\centering
\begin{tabular}{|l|llllll|}
\hline
\textbf{Topic}     & \multicolumn{6}{l|}{\textbf{Keywords and importance}}              \\ \hline
\multirow{2}{*}{1} & rain         & wind     & temp     & disaster  & humidity & weather \\
                   & 0.133        & 0.039    & 0.025    & 0.022     & 0.015    & 0.011   \\ \hline
\multirow{2}{*}{2} & hurricane    & water    & overflow & help      & time     & authority \\
                   & 0.044        & 0.034    & 0.021    & 0.019     & 0.018    & 0.018   \\ \hline
\multirow{2}{*}{3} & thunderstorm & severe   & warning  & storm     & tornado  & county  \\
                   & 0.053        & 0.046    & 0.044    & 0.032     & 0.019    & 0.018   \\ \hline
\end{tabular}
\end{table}

Table~\ref{tab:doc_weights} presents some examples of tweet-topic distribution.

\begin{table}[!htbp]
\caption{Example of matching tweets to topics}
\label{tab:doc_weights}
\centering
\begin{tabular}{l|c|c|c|c|c|c|}
\cline{2-7}
                                                                                                                                                                                                                                                                                                     & \multicolumn{6}{c|}{\textbf{Topic}}                                         \\ \hline
\multicolumn{1}{|l|}{\textbf{Tweet}}                                                                                                                                                                                                                                                                 & \textbf{1} & \textbf{2} & \textbf{3} & \textbf{4} & \textbf{5} & \textbf{6} \\ \hline
\multicolumn{1}{|l|}{\begin{tabular}[c]{@{}l@{}}After a hurricane, a guy found this pittie on\\  the roof of a submerged car waiting for help\end{tabular}}                                                                                                                                          & 0.02       & 0.02       & 0.02       & 0.02       & 0.90       & 0.021      \\ \hline
\multicolumn{1}{|l|}{\begin{tabular}[c]{@{}l@{}}Happening now: A 33,000 litres of truck laden\\ with diesel was prevented from causing another\\ national disaster by Officers and men of Federal\\ Fire Service at Ojuelegba bridge. Situation under\\ control and normalcy restored.\end{tabular}} & 0.0        & 0.0        & 0.0        & 0.96       & 0.0        & 0.0        \\ \hline
\end{tabular}
\end{table}

\subsubsection{Use case 2: COVID-19 Infection Hazards.}
\begin{figure}[!htb]
    \centering
    \subfloat[Communities\label{fig:communities}]{{\includegraphics[width=0.42\columnwidth]{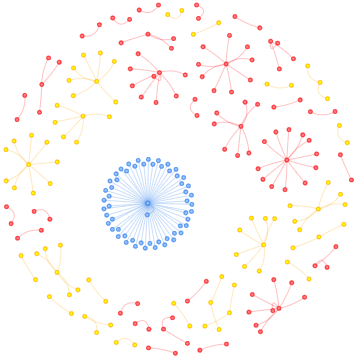}}}
    \hfill    
    \subfloat[Word cloud for "blue" community\label{fig:wordcloud}]{{\includegraphics[width=0.57\columnwidth]{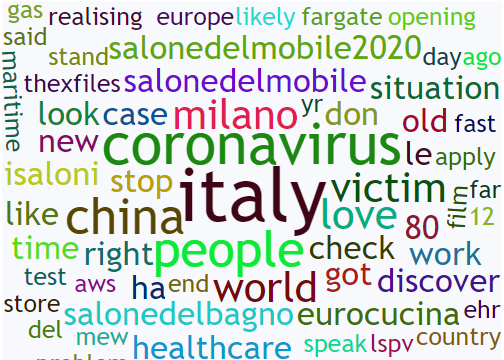}}}
    \caption{COVID-19 Infection Hazards}
    \label{fig:covid}
\end{figure}

As in the first use case, we train an OLDA model to create geolocation-content based communities on COVID-19 related topics.
Figure~\ref{fig:communities} depicts the communities for three major Coronavirus related topics, while Figure~\ref{fig:wordcloud} shows the word cloud for the "blue" topic from Figure~\ref{fig:communities} (\#1 in Table~\ref{tab:lda_covid}).
We extract 3 topics (Table~\ref{tab:lda_covid}) for which we obtain a perplexity score of $9.01$ and a $C_V$ score of $4.8$ for the entire dataset.
We obtain the following score for the community graphs: $0.0197$ Davies-Bouldin, $\sim 2.62e9$  Calinski-Harabasz, and $0.9853$ Silhouette.
By evaluating these scores, we observe that we obtain well separated and bounded clusters.

\begin{table}[!htbp]
\caption{Example of OLDA topics for COVID-19 hazards}
\label{tab:lda_covid}
\centering
\begin{tabular}{|l|llllll|}
\hline
\textbf{Topic}     & \multicolumn{6}{l|}{\textbf{Keywords and importance}}              \\ \hline
\multirow{2}{*}{1} & cases        & home     & intensive  & care     & \#coronavirus & quarantine \\
                   & 0.090        & 0.080    & 0.080      & 0.079    & 0.072        & 0.070   \\ \hline
\multirow{2}{*}{2} & coronavirus  & covid-19 & hospital   & virus    & wuhan        & chinese   \\
                   & 0.116        & 0.078    & 0.035      & 0.035    & 0.035        & 0.034   \\ \hline
\multirow{2}{*}{3} & \#coronavirus & outbreak & health     & vaccine  & spread       & died  \\
                   & 0.198        & 0.119    & 0.090      & 0.080    & 0.072        & 0.069   \\ \hline
\end{tabular}
\end{table}

\subsection{Scalability}

To test the scalability of the proposed system, we simulate multiple scenarios that simultaneously use two types of processes in a pseudo-distributed environment. 
The first type of process connects to the MongoDB database, inserts a new tweet in the \textit{tweets} collections, and updates its topic-dependent collection.
The second type of process simulates multiple clients that count all the tweets present at a given moment for a topic of interest.
This set of experiments are run on an IBM System x3550 M4 with 64GB of RAM and an Intel(R) Xeon(R) CPU E5-2670 v2 @ 2.50GHz with 40 cores.

Table~\ref{tab:scalability} presents the results for 10 executions.
For a tiny number of write and read operations, i.e., the number of tweets writes $\le$ 1\,000 and the number of clients reads $\le$ 1\,000, the system is stable and handles the burst of requests in under a second.
In this case, the maximum response time of 0.55 seconds is registered for 1\,000 write and 1\,000 read operations. 
As the number of operations increases, the system handles the request in under 1.5 minutes.
We note that the probability of receiving a high number of tweets and having a large number of client requests, even during a disaster, is very low.

\begin{table}[!htbp]
\caption{Runtime in seconds: write vs. read operations}
\label{tab:scalability}
\centering
\begin{tabular}{lr|r|r|r|r|}
\cline{3-6}
                                                           & \multicolumn{1}{l|}{} & \multicolumn{4}{c|}{\textbf{No. Clients}}                                  \\ \cline{3-6} 
                                                           &                       & \multicolumn{1}{c|}{\textbf{100}} & \multicolumn{1}{c|}{\textbf{1\,000}} & \multicolumn{1}{c|}{\textbf{10\,000}} & \multicolumn{1}{c|}{\textbf{100\,000}} \\ \hline
\multicolumn{1}{|l|}{\parbox[t]{1mm}{\multirow{4}{*}{\rotatebox[origin=c]{90}{\tiny \textbf{No. Tweets}}}}} & \textbf{100}          &  0.13 $\pm$ 0.01 &  0.54 $\pm$ 0.01 &  4.49 $\pm$ 0.16 & 44.10 $\pm$ 0.30  \\ \cline{2-6} 
\multicolumn{1}{|l|}{}                                     & \textbf{1\,000}         &  0.36 $\pm$ 0.01 &  0.55 $\pm$ 0.01 &  4.43 $\pm$ 0.05 & 43.67 $\pm$ 1.21  \\ \cline{2-6} 
\multicolumn{1}{|l|}{}                                     & \textbf{10\,000}        &  3.45 $\pm$ 0.11 &  3.59 $\pm$ 0.03 &  4.80 $\pm$ 0.02 & 48.05 $\pm$ 0.05  \\ \cline{2-6} 
\multicolumn{1}{|l|}{}                                     & \textbf{100\,000}       & 33.37 $\pm$ 0.48 & 34.60 $\pm$ 0.07 & 34.83 $\pm$ 0.77 & 87.67 $\pm$ 1.04 \\ \hline
\end{tabular}
\end{table}

\section{Conclusions}\label{sec:conclusions}

In this paper, we present ContCommRTD, a new distributed system that determines geolocation-content based communities depending on the topics of interest and user geolocation, and takes into account misinformation on social networks.
Moreover, we have also shown how our system can be applied to track the evolution of different hazards.
This is of significant importance, especially for government organizations that can subscribe to topics of interest in a specific geographical area or can subscribe to receive information on a topic regardless of the location. 
Our system also provides an interactive graphical interface where a user can select what are the topics of interest.
On the user map, the notifications within the relevant topics will appear colored.
All the notifications that came from the detection of other information apart from the chosen topics are displayed in gray color. 
Furthermore, our approach can be generalized for other types of hazards or social events if new terms are added to the dictionary used to collect specific tweets.
As tweets collected from unreliable or unchecked sources may spread misinformation, we consider detecting these tweets and removing them from our active communities. 
For this task, we propose FN-BERT-TFIDF, a new Deep Learning BERT-based model.
Moreover, the proposed system scales well as the number of read and write operations increases.

As future work, we plan to add more information in order to analyze human behaviors during disasters, as follows.
Between the filtering and preprocessing process, the tweet could undergo a more detailed analysis, such as domain-specific rule-based filtering and disambiguation, and text-based sentiment analysis~\cite{Lukasiewicz2014,Xu2021,Truica2021}.

\bibliographystyle{plainnat}  
\bibliography{main}

\end{document}